\documentclass{article} 
\usepackage{amsmath,amssymb,graphicx}
\usepackage{url}
\usepackage{apacite}
\usepackage[english]{babel}
\usepackage[top=1.5in, bottom=1.5in, left=1.5in, right=1.5in]{geometry}

\title{Computing Exact Clustering Posteriors \\ with Subset Convolution}

\author{
Jukka Kohonen\thanks{Corresponding author}\\
Department of Mathematics and Statistics\\
P.O. Box 68\\
FI-00014 University of Helsinki\\
\texttt{jukka.kohonen@helsinki.fi} \\
\and
Jukka Corander\\
Department of Mathematics and Statistics\\
P.O. Box 68\\
FI-00014 University of Helsinki\\
\texttt{jukka.corander@helsinki.fi} \\
\\
Short title: Exact clustering with subset convolution 
}

\newcommand\given{\;|\;}
\DeclareMathOperator{\Gammadist}{Gamma}
\DeclareMathOperator{\Bernoullidist}{Bernoulli}
\DeclareMathOperator{\Betadist}{Beta}
\newtheorem{proposition1}{Proposition}
\newtheorem{corollary1}{Corollary}

\begin{document}

\maketitle 

\begin{abstract}
An exponential-time exact algorithm is provided for the task of
clustering $n$ items of data into $k$ clusters.  Instead of seeking
one partition, posterior probabilities are computed for summary
statistics: the number of clusters, and pairwise co-occurrence.  The
method is based on subset convolution, and yields the posterior
distribution for the number of clusters in $O(n3^n)$ operations, or
$O(n^32^n)$ using fast subset convolution.  Pairwise co-occurrence
probabilities are then obtained in $O(n^32^n)$ operations.  This is
considerably faster than exhaustive enumeration of all partitions.
\end{abstract}


\section{Introduction}
The vast majority of clustering literature is dedicated to finding one
particularly good partition, i.e. a definite clustering of the data.
In a probabilistic setting, a good partition may be defined as one
that has high likelihood, or high posterior probability compared to
other partitions that have been considered.  However, posterior
probabilities are usually known only up to an unknown normalizing
constant over the clustering space.  Thus, one may deduce that one
partition is, say, $10^{20}$ times more probable than another
partition, while having no idea whether the posterior probability
itself is on the order of $0.5$, or perhaps $10^{-10}$.  Clearly, from
the perspective of Bayesian inference, this is an unfortunate
situation, and it is in general unknown how well standard Monte Carlo
sampling strategies \cite<e.g.,>{neal2000,jain2004,huelsenbeck2007}
can approximate the true partition posterior.

Furthermore, even the optimal partition may have a vanishingly tiny
posterior probability. For instance, if the clusters are not very
clearly distinguishable in the data, the posterior mass may be spread
over a large number of partitions. Suppose that the optimal partition
has $k=4$ clusters and the posterior probability $10^{-10}$. It
appears reasonable to claim that posterior inferences should also be
concerned with the remaining posterior mass $1 - 10^{-10} \approx 1$,
and in particular how the probability mass is spread over different
values of $k$.  In addition, it remains in practice unknown how the
posterior distribution over possible data partitions is affected by
the dimensionality of observed features, as well as by the choice of a
model and prior distribution for model parameters and partitions, due
to the rapidly increasing size of the clustering space, which makes
full enumeration infeasible in practice.

Since posterior probabilities would be highly desirable for meaningful
{\em collections} of partitions, we introduce here an approach to
their efficient calculation based on subset convolution.  In
particular, we are interested in posterior probabilities for the
following two kinds of propositions: (1) ``The data are appropriately
represented by $k$ distinct clusters.''  (2) ``The two items $i$ and
$j$ belong to the same cluster'' (for each pair $i,j$).  We shall show
how these probabilities can be exactly evaluated with subset
convolution without actually enumerating all partitions. The pairwise
co-occurrence probabilities in the latter proposition are also
directly useful for deriving a model-averaged estimate of the
partition under specific loss-functions and have been considered by
multiple authors
\cite<e.g.,>{dawson2001,huelsenbeck2007,lau2007,corander2009}.

To demonstrate the use of a subset convolution, we consider a variant
of the {\em product partition model}, which has been studied, \citeA<e.g.,
by>{hartigan1990,barry1992,quintana2003,lau2007,corander2009,dahl2009}.

A dynamic programming method to find the optimal partition was first
suggested by \citeA{jensen1969}, and implemented by
\citeA{hubert2001}.  \citeA{os2004} proposed various speedups to the
original method, but in general the dynamic programming approach has
time requirement $O(3^n)$.  In a line of different work,
\citeA{dahl2009} showed how dynamic programming can be used to
efficiently find the posterior mode partition even for very large sets
of items, however, the method is restricted only to the case where
sufficient statistics from data are univariate for each cluster. While
the goal of searching for optimum is very different from computing the
posterior, computationally they involve highly similar steps.

Applications of subset convolution to various combinatorial problems,
including partitioning, are described by \citeA{bjorklund2007} and
\citeA{fomin2010}.  To the best of our knowledge, use of subset
convolution to computing posteriors of $k$ and pairwise co-occurrence
has not been considered previously.

The remainder of the paper is structured as follows. The main results
are derived in the three subsequent sections and some numerical
illustrations are given in the penultimate section.The final section
concludes with some remarks and discussion about potential
generalizations and wider application of the presented ideas.


\section{Definitions and preliminaries}
Let $U$ denote a set of $n$ elements, or {\em items}, labeled by
integers $\{1,\ldots,n\}$.  Each data item $i$ is associated with some
$D$-dimensional feature vector $y_i = (y_{i1},\ldots,y_{iD})$, and the
whole data set will be denoted by $y$.

A {\em cluster} is a subset of $U$.  An {\em unordered partition} of
$U$ is a set of disjoint, nonempty clusters whose union is $U$.  An
{\em ordered partition} is a tuple of disjoint, nonempty clusters
whose union is $U$.  A partition of cardinality $k$ is called a {\em
  $k$-partition}.  In a {\em singleton partition} each item forms its
own cluster ($k=n$).  In a {\em trivial partition} all items are
clustered together ($k=1$).

The distinction between ordered and unordered partitions is crucial
when counting partitions, computing sums, or defining probability
distributions over them.  The distinction is also at the roots of the
so-called label switching problem, discussed e.g. by
\citeA{stephens2000}.  The number of unordered $k$-partitions of $n$
items is the Stirling number of the second kind, denoted $S(n,k)$,
while the number of ordered $k$-partitions is $k!  \times S(n,k)$.
Consider the intuitive notion of ``the'' singleton partition: it is
either unique (unordered) or there are $n!$ of them (ordered).  The
trivial partition, in contrast, is unique in both cases.  In the
following, a partition is assumed to be unordered unless otherwise
noted.

The task of clustering in general is to characterize particularly good
or plausible data partitions in some statistical sense.  We adopt here
the Bayesian perspective, where the (prior) predictive probability of
the data (also termed as {\em evidence}) is conditioned on the
partition, and seek to characterize posterior probability within the
space of possible partitions.  This is in general a daunting task
since the partition space grows quickly with respect to $n$.  For
instance, suppose $n=20$, then, the number of 4-partitions alone is
$S(20,4) \approx 4.5 \times 10^{10}$, and the number of all partitions
(for $k=1,\ldots,20$) is the 20th Bell number, about $5.2 \times
10^{13}$.  A brute force search or summation over them would be a
considerable computing task and similarly, any practically obtainable
Monte Carlo sample from the posterior will only cover a small fraction
of the space.


\section{Partition posterior and subset convolution}
\newcommand{\fprior}{f_\text{prior}}
\newcommand{\flik}{f_\text{lik}}

Our method targets posterior distribution under a {\em product
  partition model}
\cite{hartigan1990,barry1992,quintana2003,dahl2009}, extended to
accommodate an arbitrary prior for $k$, the number of clusters.  We
assume that the prior probability for an ordered $k$-partition $S =
(S_1,\ldots,S_k)$ factorizes as
\begin{equation*}
p(S) = w_k \cdot \prod_{j=1}^k \fprior(S_j), 
\end{equation*}
where $\fprior$ is an arbitrary function defined for the nonempty
subsets of $U$, and the factors $w_1,\ldots,w_n$ control the marginal
prior probability for $k$.  Likewise, we assume that partition
marginal likelihood (evidence) factorizes as
$$
p(y \given S) = \prod_{j=1}^k p(y_{(j)} \given S_j) = \prod_{j=1}^k \flik(S_j),
$$ where $\flik$ expresses the marginal likelihood of data $y_{(j)}$
observed in cluster $S_j$, and is also an arbitrary function over
subsets.  We shall later show examples of standard models satisfying
these desiderata. However, our approach could also be applied in the
case where marginal likelihoods are not analytically available but are
replaced with approximations, such as those based on the Laplace
method.

For simplicity $\fprior$ and $\flik$ are combined as a single function
$f(X) = \fprior(X) \flik(X)$.  For completeness we define
$f(\varnothing)=0$, which rules out empty clusters.  This function $f$
and the factors $w_1,\ldots,w_n$ are the input to our clustering
model.

The posterior probability of an ordered $k$-partition is now
\begin{equation*}
p(S \given y)
 = \frac{p(S) p(y \given S)}{p(y)}
 = Z \cdot w_k \cdot \prod_{j=1}^k f(S_j),
\end{equation*}
where $Z=1/p(y)$ is the normalizing constant.

Note that we define the model for {\em ordered} partitions for
computational convenience.  In practice each unordered $k$-partition
is represented as $k!$ ordered $k$-partitions due to permutation.

This model accommodates various partition priors.  A widely used
prior, where all unordered partitions are equiprobable ({\em uniform
  on partitions}), is obtained by setting $\fprior(X)=1$ for all $X
\ne \varnothing$, and $w_k = 1/(k! \cdot B_n)$, where $B_n$ is the
$n$th Bell number.  Under this prior, the marginal distribution for
$k$ is highly nonuniform.

Another prior is {\em uniform on k}, where $p(k)=1/n$ for
$k=1,\ldots,n$, and partitions of the same cardinality are
equiprobable.  This prior, obtained by setting $\fprior(X)=1$ and $w_k
= 1/(k!  \cdot n \cdot S(n,k))$, is a convenient way of expressing no
strong prior belief about $k$.  It does not seem widely used, but
occurs as a special case of a larger prior family introduced by
\citeA{knorrheld2000}.  \citeA{quintana2003} prove that this
prior cannot be expressed as an ordinary product partition model.

Yet another prior is based on a Dirichlet process (DP) with weight
parameter $\theta$ \cite{neal2000,quintana2003}.  This is obtained by
$\fprior(S_j) = \Gamma(|S_j|)$ and $w_k =
\Gamma(\theta)/(\Gamma(\theta+n) \cdot k!)$.

\subsection{Computing posterior of $k$}
\label{sec:pk}
\newcommand{\Sspace}{{\cal S}}
\newcommand{\sumk}{\sum_{\substack{S \in \Sspace \\ |S|=k}}}
\newcommand{\prodk}{{\prod_{j=1}^k}}
The posterior probability of $k$ clusters equals under the above formulation
\begin{equation}
p(k \given y) = Z \cdot w_k \cdot \sumk \prodk f(S_j),
\label{pky}
\end{equation}
where $\Sspace$ is the space of ordered partitions.  This sum of
products is conveniently expressed by means of subset convolution.
Given two real-valued functions $f$ and $g$ defined on the subsets of
$U$, their {\em subset convolution}, or {\em convolution} for short,
is the function
\begin{equation}
(f*g)(X) = \sum_{A \subseteq X} f(A) \cdot g(X \setminus A),
 \quad\text{for all $X \subseteq U$},
\label{con}
\end{equation}
or equivalently in a more symmetric form,
\begin{equation*}
(f*g)(X) = \sum_{\substack{A,B \subseteq X\\ A+B=X}} f(A) \cdot g(B),
\end{equation*}
where $A+B=X$ represents disjoint union.  Convolution is
associative, and iterative application yields
$$
(f_1 * \ldots * f_k)(X) =
\sum_{\substack{A_1,\ldots,A_k \subseteq X\\ A_1+\ldots+A_k=X}} \prod_{j=1}^k f_j(A_j).
$$
In other words, $k$-fold convolution expresses summation over ordered
$k$-partitions of a set $X$.  Writing \eqref{pky} in terms of iterated
convolution, we arrive at the following proposition.

\begin{proposition1}[Posterior of $k$]
The posterior probability for the number of clusters being $k$
is
\begin{equation}
p(k \given y) = Z \cdot w_k \cdot f^{(k)}(U),
\label{pkyconvolution}
\end{equation}
where $f^{(k)} = (f * \ldots * f)$ denotes the convolution of $k$
copies of $f$.
\end{proposition1}
The normalizing constant $Z$ is obtained from the constraint
$\sum_{k=1}^n p(k \given y) = 1$.

For a single set $X$, a direct evaluation of the sum \eqref{con}
yields the convolution $(f*g)(X)$ in $O(2^{|X|})$ arithmetic
operations.  Repeating for all $X \subseteq U$ yields the full
convolution table for $f*g$ in $O(3^n)$ operations; we shall call this
method the {\em direct subset convolution}.  Assuming that $f$ has
been tabulated for all $2^n$ subsets of $U$ (all possible clusters),
the tables for $f^{(2)}, \ldots, f^{(n)}$ can be computed iteratively
with $n-1$ convolutions.  Since the exact posterior distribution for
$k$ can be computed from the quantities $f(U), f^{(2)}(U), \ldots,
f^{(n)}(U)$ using \eqref{pkyconvolution}, we obtain the following.

\begin{corollary1}
If $f$ has been computed for all $X \subseteq U$, the full posterior
distribution for $k$ can be computed in $O(n 3^n)$ arithmetic
operations.
\end{corollary1}

For example, for $n=20$ items, the number of operations needed
is on the order of $n 3^n \approx 7 \times 10^{10}$, much less than
if the $5.2 \times 10^{13}$ possible (unordered) partitions were
actually enumerated and their posterior probabilities computed
one by one.

A further speedup for large $n$ can be achieved by using {\em fast
  subset convolution}, invented by \citeA{bjorklund2007}, which
requires $O(n^2 2^n)$ operations instead of $O(3^n)$.  However, for
moderate values of $n$, such as 20, the savings are not extensive.  An
additional complication is that fast subset convolution involves both
addition and subtraction, potentially leading to large rounding errors
in floating point arithmetic.  In our experiments, already for $n=18$
rounding errors caused the result from fast subset convolution to be
completely erroneous.  This can be avoided, with some extra
computational cost, by using exact arithmetic with arbitrary-precision
integers with a software library such as GMP \cite{GMP}.  In
comparison, direct subset convolution does not involve subtraction,
and in our experiments floating point arithmetic has been sufficiently
accurate.

\subsection{Computing posterior pairwise co-occurrence}
\label{sec:ppair}
The main goal of clustering is to identify which items belong together
and which don't.  By asking this question for all items
simultaneously, one is led to seeking a single partition as the
candidate for how items should be merged or separated from each other.
But in numerous situations the data do not clearly distinguish one
partition as the correct one, and many alternative partitions may have
considerable posterior mass.  For meaningful posterior conclusions the
partition probabilities need to be summarized in a sensible manner to
provide model-averaged inference.

A natural approach is to consider each pair $i,j$ of items at a time
and evaluate the posterior probability of the event $C_{ij}$ that they
belong to the same cluster.  We shall call this the posterior {\em
  pairwise co-occurrence} (probability).  If posterior pairwise
co-occurrence is computed for all item pairs, the results can be
summarized as a {\em co-occurrence matrix}, as suggested by 
\citeA{ohagan1997} and also considered more in detail by
\citeA{dawson2001,huelsenbeck2007,lau2007,corander2009}. In particular,
it can be shown that an optimal partition estimate can be derived from
the co-occurrence matrix under a more natural loss function than the
zero-one loss leading to the choice of mode partition
\cite{corander2009}.  In \cite{lau2007}, a partition is sought that
minimizes expected loss, where loss is defined by the numbers of
misassigned item pairs.  The expected loss, over the space of
partitions, can be directly computed from the co-occurrence matrix, if
that is available.

Consider first the joint posterior for $(k, C_{ij})$, i.e. the
probability that the data comes from exactly $k$ clusters such that
items $i$ and $j$ are in the same cluster. This is obtained by summing
the posterior \eqref{pky} over all ordered $k$-partitions where
$C_{ij}$ holds.  Now, since under our assumptions likelihood and
partition prior are symmetric with respect to cluster indexing, all
permutations of an ordered partition have the same probability; in
particular, the probability that items $i$ and $j$ are in the same
cluster (one of $S_1,\ldots,S_k$) equals $k$ times the probability
that they are in the first cluster $S_1$.  Since $U \setminus S_1$
must be covered by the other clusters $S_2,\ldots,S_n$, we have
$$
p(k, C_{ij} \given y) = 
  Z \cdot k \cdot w_k \cdot
  \sum_{\substack{S_1 \subseteq U \\ i,j \in S_1}}
  f(S_1) \cdot (f^{(k-1)})(U \setminus S_1).
$$
Summing over partition cardinalities we obtain the following
proposition.

\begin{proposition1}[Posterior pairwise probability]
The posterior probability for items $i$ and $j$ being in the
same cluster equals
\begin{equation}
p(C_{ij} \given y) = Z \cdot \sum_{k=1}^n \left(
  k \cdot w_k \cdot
  \sum_{\substack{S_1 \subseteq U \\ i,j \in S_1}}
  f(S_1) \cdot (f^{(k-1)})(U \setminus S_1) \right).
\label{pairwise}
\end{equation}
\end{proposition1}

The inner sum in \eqref{pairwise} has $2^{n-2}$ terms.  Repeating for
all pairs $i,j$ we have the following:

\begin{corollary1}
If the iterated convolutions $f^{(2)},\ldots,f^{(n-1)}$ have been
computed, the full posterior co-occurrence matrix can be computed in
$O(n^32^n)$ arithmetic operations.
\end{corollary1}

\subsection{Finding the mode partition}
\label{sec:mode}
Although our emphasis lies in posterior summary statistics over
partitions, it is worth noting that a slight variant of subset
convolution can be used for finding the mode partition.  If the
summation in \eqref{pky} is replaced with maximization, one obtains
the maximum posterior probability among $k$-partitions.  Now this can
be computed using a variant of subset convolution, where the summation
is replaced with maximization (i.e., the subset convolution is
performed over the max-product semiring, instead of the usual
sum-product ring).  This yields an $O(n 3^n)$ algorithm for finding
the maximum-probability $k$-partitions for $k=1,\ldots,n$.  The
maximum among those is of course the global mode partition.  In fact,
this method is equivalent to Jensen's dynamic programming algorithm
\cite{jensen1969}, now expressed in terms of subset convolution.


\section{Examples of data models}
Our method takes as its input a table of the function $f(X)$ for all
$2^n-1$ nonempty subsets of $U$.  Thus no restrictions are placed on
the form of the likelihood function, as long as it can be feasibly
computed for $2^n-1$ sets.  We have experimented with two models where
the marginal likelihood is analytically available.  It should be noted
that the general method of subset convolution is not limited to these
two models.  For example, for discrete data the beta-binomial model
generalizes in a straightforward fashion to a Dirichlet-multinomial or
gamma-Poisson family of distributions.

\subsection{Beta-binomial model for binary data}
For binary data we assume a Bernoulli distribution with a beta prior
\cite[pp. 157, 160]{degroot1970}.  For each cluster $S_j$ and feature
$d$, independently from other clusters and features, we assume an
unknown parameter $m_{jd}$ (cluster mean) such that
\begin{align*}
m_{jd}                &\sim \Betadist(\alpha,\beta) \\
(y_{id} \given m_{jd}) &\sim \Bernoullidist(m_{jd}),
   \quad\text{for items $i \in S_j$},
\end{align*}
where $\alpha,\beta$ are prior hyperparameters.  This implies that
within a cluster and a variable, the counts of zeros and ones are
binomially distributed, conditional on $m_{jd}$.  The marginal
likelihood for the data $y_{(j)d}$ observed in cluster $S_j$ for
feature $d$ can be expressed in terms of the sufficient statistics
$(c,s)$, where $c=|S_j|$ is cluster size (number of items), and
$s=\sum_{i \in S_j} y_{ji}$ is the number of observed ones.
Integrating out the binomial parameters, we obtain the marginal
likelihood
$$
p(y_{(j)d}) = \frac{\Gamma(\alpha+s) \Gamma(\beta+c-s) \Gamma(\alpha+\beta)}
{\Gamma(\alpha+\beta+c) \Gamma(\alpha) \Gamma(\beta)}.
$$

\subsection{Gamma-normal model for continuous data}
For continuous data we assume normal distribution with a normal-gamma
prior, as described by, e.g., \citeA[pp. 168--171]{degroot1970} and
\citeA[p. 440]{bernardo1994}.  For each cluster $S_j$ and feature $d$,
we assume two unknown parameters $m_{jd}$ (cluster mean) and $r_{jd}$
(cluster precision) such that
\begin{align*}
r_{jd}                        &\sim \Gammadist(\alpha,\beta)     \\
(m_{jd} \given r_{jd})         &\sim N(\mu, 1/(\tau r_{jd})) \\
(y_{id} \given m_{jd},r_{jd})   &\sim N(m_{jd}, 1/r_{jd}),
  \quad\text{for items $i \in S_j$},
\end{align*}
where $\alpha,\beta,\mu,\tau$ are prior hyperparameters.  The marginal
likelihood can be expressed in terms of the sufficient statistics
$(c,s,q)$, where $c = |S_j|$ is cluster size, $s = \sum_{i \in S_j}
y_{ji}$ is the sum of data, and $q = \sum_{i \in S_j} y_{ji}^2$ is the
sum of squared data.  The marginal likelihood is derived e.g. by
\citeA{murphy2007}, and in our notation it becomes
$$
p(y_{(j)d}) =
  \frac{\Gamma(\alpha_c)}{\Gamma(\alpha)}
  \frac{\beta^\alpha}{\beta_c^{\alpha_c}}
  \left( \frac{\tau}{\tau_c} \right) ^ {1/2}
  (2\pi) ^ {-c/2},
$$
where
\begin{align*}
  \alpha_c &= \alpha + c/2,   \\
  \beta_c  &= \beta  + \frac{q - s^2/c}{2} + \frac{\tau (s-c\mu)^2/c}{2(\tau+c)},\\
  \tau_c   &= \tau   + c.
\end{align*}


\section{Experiments}

\subsection{Clustering posteriors with a continuous model}
\begin{figure}[b!]
  \includegraphics[width=4.5cm]{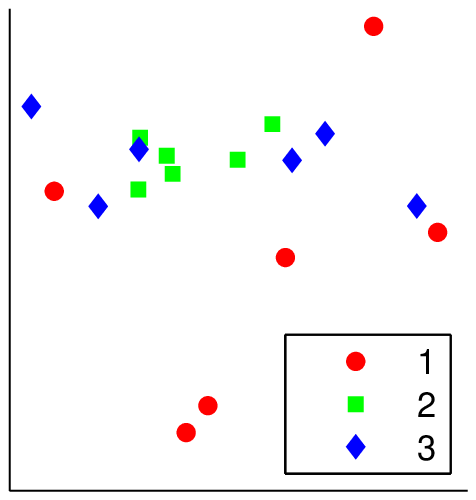}
  \hfill
  \includegraphics[width=4.5cm]{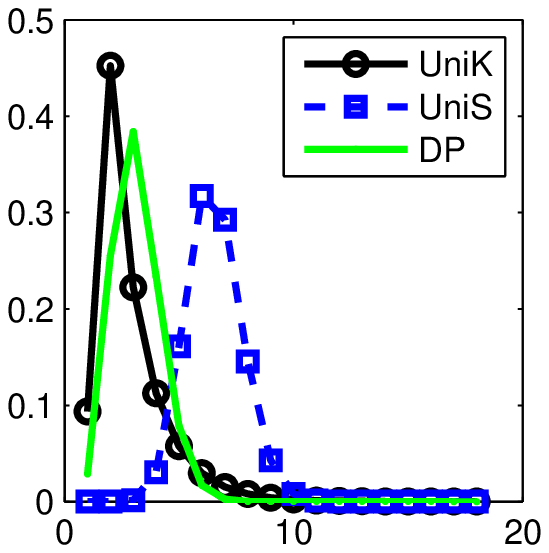}
  \hfill
  \includegraphics[width=4.5cm]{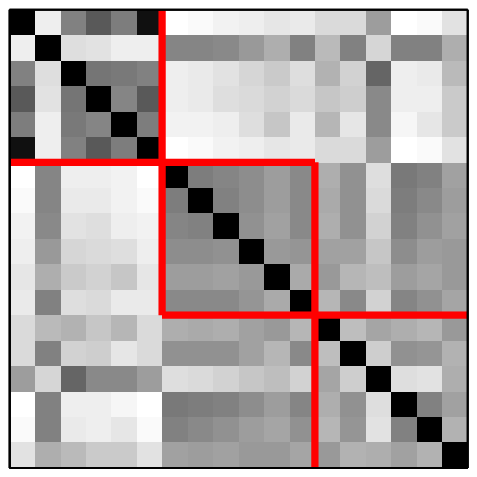}

  \caption{Simulated normal data, $k=3$, $n=18$, $D=2$.  Left:
    original data.  Center: Posterior distribution for $k$ with three
    priors (uniform on $k$, uniform on partitions, and DP with
    $\theta=1$).  Right: Posterior pairwise co-occurrence matrix (dark
    indicates high probability; red lines highlight the generating
    clusters).}  \label{fig1}
\end{figure}

For a simple illustration, let us consider $18$ items generated from
$3$ clusters of 6 items each, with bivariate normal data, where the
cluster parameters $(m_{jd},r_{jd})$ were randomly generated with
hyperparameters $\alpha=1$, $\beta=1$, $\mu=0$, $\tau=1$.  The data
are shown in Figure~\ref{fig1} (left), colored by the true
(generating) partition.

Assuming the hyperparameters known, but partition unknown, the
posteriors for $k$ and pairwise co-occurrence were computed using
subset convolution.  Assuming uniform prior on $k$, the posterior
(Figure~\ref{fig1} center, black line) is peaked at $k=2$, which
seems reasonable by visual inspection of the data, as the clusters 2
and 3 overlap considerably.  The posterior distribution shows the
inherent uncertainty over $k$; computing just the mode partition would
not provide such information.

With a DP prior ($\theta=1$), the posterior is similar, but peaked at
$k=3$; note that the prior is itself peaked at $k=3$, and in general
favors partitions of small cardinality.

If all unordered partitions are assumed {\em a priori} equiprobable,
the posterior (Figure~\ref{fig1} center, blue line) is peaked at
$k=6$.  This undesirable behavior is due to the strong prior
preference for large partitions, simply because there are so many of
them.  For example, there are $S(18,3) \approx 6.4\times 10^7$
unordered $3$-partitions, but $S(18,6) \approx 1.1\times 10^{11}$
unordered $6$-partitions.  Assuming them equiprobable implies a prior
belief that $k=6$ is about $1700$ times more probable than $k=3$.

The matrix of posterior pairwise co-occurrence probabilities (assuming
uniform prior on $k$) is shown in Figure~\ref{fig1} (right).  The
items are ordered according to the generating partition for visual
inspection.  The first cluster stands clearly apart (with the
exception of the second item, which is the red dot on the far left).

We also computed the optimal partitions for $k=1,\ldots,n$ using the
method described in subsection~\ref{sec:mode}, again assuming uniform
prior on $k$.  The global mode turns out to be the trivial partition
with posterior probability $0.094$.  For $k=2,3$ the optimal
partitions have much lower posterior probabilities $0.014$ and
$0.001$, respectively.  It is apparent that the mode partition in
itself is not well representative of the full posterior distribution.

The computation of the full posterior for $k$ and pairwise
co-occurrence takes about 7~minutes of CPU time on a 2.4~GHz AMD
Opteron, using a C implementation of direct subset convolution.  We
estimate that full enumeration of all unordered partitions would have
taken about 64~hours.


\subsection{Clustering posteriors with a binary model}
\begin{figure}[b!]
  \includegraphics[width=4.5cm]{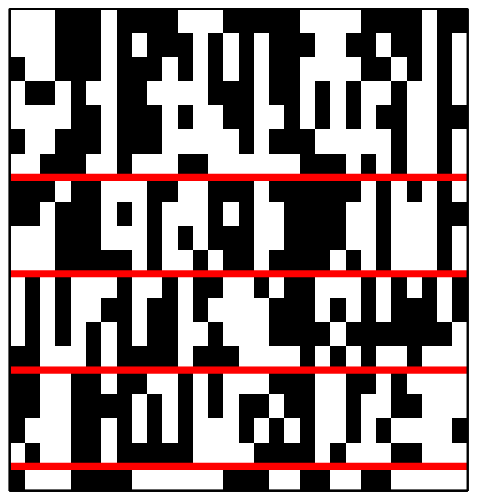}
  \hfill
  \includegraphics[width=4.5cm]{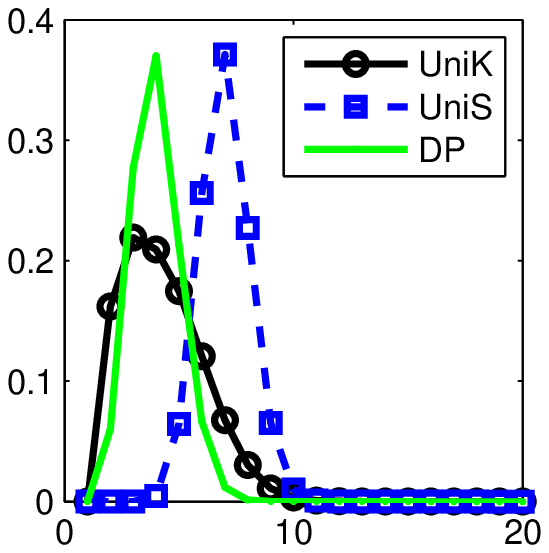}
  \hfill
  \includegraphics[width=4.5cm]{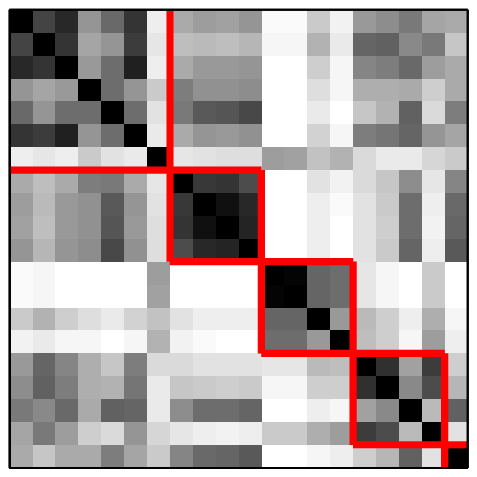}

  \caption{Simulated binary data, $k=5$, $n=20$, $D=30$.  Left:
    original data (red lines indicate cluster boundaries).  Center:
    Posterior distribution for $k$ with three priors (uniform on $k$,
    uniform on partitions, and DP with $\theta=1$).  Right: Posterior
    pairwise co-occurrence matrix (dark indicates high probability;
    red lines highlight the generating clusters.}  \label{fig2}
\end{figure}
Figure~\ref{fig2} illustrates an experiment with simulated data
from five clusters, with 20 items and 30 binary features.  The
generating partition is a randomly chosen 5-partition and the data
were generated with hyperparameters $\alpha=\beta=1$.

The posterior distributions for $k$ (Figure~\ref{fig2} center) and for
pairwise co-occurrence (Figure~\ref{fig2} right) show the inherent
uncertainty over the correct partition.  Yet they also indicate
summary statistics that can be reasonably estimated.  For example, the
pairwise matrix indicates that the four items 8--11 probably belong
together (which is correct according to the generating model);
likewise for items 12--15.

The mode partition has $k=2$ and posterior probability $0.0267$.  For
$k=3,4,5$ the optimal partitions have probabilities $0.0050$, $0.0026$
and $0.0005$, respectively.  Judging from these posterior
probabilities alone --- especially if they were unnormalized, and only
their ratios could be seen --- one might deduce that $k=2$ is an
overwhelmingly good model for the data, and that $k=5$ is quite
unlikely.  Yet the summary statistics lend considerable support to the
possibility that $k=5$ (which corresponds to the generating model).
Again we must note that posterior probabilities of single partitions
do not well represent the full posterior distribution.

The shown posterior distributions are exact, and all uncertainty is
due to the data and the probability model assumed; not due to any
computational approximation.  If the data were more informative, the
posterior distributions would correspondingly be more peaked.  Here
computing the posteriors for $n=20$ took about 3~hours of CPU time
whereas a full enumeration of the partitions would take approximately
200~days.


\section{Discussion}

The convolution approach introduced here has potential for multiple
purposes in cluster analysis.  For instance, by enabling an exact
evaluation of the posterior over the number of clusters and pairwise
co-occurrence probabilities, one can investigate how the
dimensionality of the feature space and choices of prior
hyperparameters affect the power to detect clusters in a particular
modeling scenario. Another application is to use the exact posterior
in a proposal operator for an Markov chain Monte Carlo sampling
algorithm.  Obviously, the exponential time requirement limits the
applicability of the method to fairly modest instances, on the order
of 20--25 items.  Even so, we think that having an exact posterior at
least in such cases can serve as a useful ``gold standard'' when
evaluating the performance and the characteristics of other, more
practical methods. Furthermore, the exact posteriors could be used for
larger data sets by segmenting the data into several small subsets and
evaluating the posteriors separately for each of them. It appears as
an attractive target for further research to investigate intelligent
strategies for combining posterior information from the different
segments and then proceeding towards global partition inferences.


\subsubsection*{Acknowledgments}
This research was funded by the ERC grant no. 239784 and AoF grant no. 251170.

\bibliographystyle{apacite}
\bibliography{refs}

\begin{thebibliography}{}

\bibitem [\protect \citeauthoryear {%
Barry%
\ \BBA {} Hartigan%
}{%
Barry%
\ \BBA {} Hartigan%
}{%
{\protect \APACyear {1992}}%
}]{%
barry1992}
\APACinsertmetastar {%
barry1992}%
\begin{APACrefauthors}%
Barry, D.%
\BCBT {}\ \BBA {} Hartigan, J.%
\end{APACrefauthors}%
\unskip\
\newblock
\APACrefYearMonthDay{1992}{}{}.
\newblock
{\BBOQ}\APACrefatitle {Product partition models for change point problems}
  {Product partition models for change point problems}.{\BBCQ}
\newblock
\APACjournalVolNumPages{The Annals of Statistics}{20}{1}{260--279}.
\PrintBackRefs{\CurrentBib}

\bibitem [\protect \citeauthoryear {%
Bernardo%
\ \BBA {} Smith%
}{%
Bernardo%
\ \BBA {} Smith%
}{%
{\protect \APACyear {1994}}%
}]{%
bernardo1994}
\APACinsertmetastar {%
bernardo1994}%
\begin{APACrefauthors}%
Bernardo, J\BPBI M.%
\BCBT {}\ \BBA {} Smith, A\BPBI F\BPBI M.%
\end{APACrefauthors}%
\unskip\
\newblock
\APACrefYear{1994}.
\newblock
\APACrefbtitle {Bayesian Theory} {Bayesian theory}.
\newblock
\APACaddressPublisher{}{John Wiley and Sons}.
\PrintBackRefs{\CurrentBib}

\bibitem [\protect \citeauthoryear {%
{Bj\"orklund}%
, Husfeldt%
, Kaski%
\BCBL {}\ \BBA {} Koivisto%
}{%
{Bj\"orklund}%
\ \protect \BOthers {.}}{%
{\protect \APACyear {2007}}%
}]{%
bjorklund2007}
\APACinsertmetastar {%
bjorklund2007}%
\begin{APACrefauthors}%
{Bj\"orklund}, A.%
, Husfeldt, T.%
, Kaski, P.%
\BCBL {}\ \BBA {} Koivisto, M.%
\end{APACrefauthors}%
\unskip\
\newblock
\APACrefYearMonthDay{2007}{}{}.
\newblock
{\BBOQ}\APACrefatitle {Fourier meets {M\"obius}: fast subset convolution}
  {Fourier meets {M\"obius}: fast subset convolution}.{\BBCQ}
\newblock
\BIn{} \APACrefbtitle {Proceedings of the thirty-ninth annual {ACM} symposium
  on {Theory} of computing ({STOC} 07)} {Proceedings of the thirty-ninth annual
  {ACM} symposium on {Theory} of computing ({STOC} 07)}\ (\BPGS\ 67--74).
\PrintBackRefs{\CurrentBib}

\bibitem [\protect \citeauthoryear {%
Corander%
, Gyllenberg%
\BCBL {}\ \BBA {} Koski%
}{%
Corander%
\ \protect \BOthers {.}}{%
{\protect \APACyear {2009}}%
}]{%
corander2009}
\APACinsertmetastar {%
corander2009}%
\begin{APACrefauthors}%
Corander, J.%
, Gyllenberg, M.%
\BCBL {}\ \BBA {} Koski, T.%
\end{APACrefauthors}%
\unskip\
\newblock
\APACrefYearMonthDay{2009}{}{}.
\newblock
{\BBOQ}\APACrefatitle {Bayesian unsupervised classification framework based on
  stochastic partitions of data and a parallel search strategy} {Bayesian
  unsupervised classification framework based on stochastic partitions of data
  and a parallel search strategy}.{\BBCQ}
\newblock
\APACjournalVolNumPages{Advances in Data Analysis and
  Classification}{3}{1}{3--24}.
\PrintBackRefs{\CurrentBib}

\bibitem [\protect \citeauthoryear {%
Dahl%
}{%
Dahl%
}{%
{\protect \APACyear {2009}}%
}]{%
dahl2009}
\APACinsertmetastar {%
dahl2009}%
\begin{APACrefauthors}%
Dahl, D.%
\end{APACrefauthors}%
\unskip\
\newblock
\APACrefYearMonthDay{2009}{}{}.
\newblock
{\BBOQ}\APACrefatitle {Modal clustering in a class of product partition models}
  {Modal clustering in a class of product partition models}.{\BBCQ}
\newblock
\APACjournalVolNumPages{Bayesian Analysis}{4}{2}{243--264}.
\PrintBackRefs{\CurrentBib}

\bibitem [\protect \citeauthoryear {%
Dawson%
\ \BBA {} Belkhir%
}{%
Dawson%
\ \BBA {} Belkhir%
}{%
{\protect \APACyear {2001}}%
}]{%
dawson2001}
\APACinsertmetastar {%
dawson2001}%
\begin{APACrefauthors}%
Dawson, K\BPBI J.%
\BCBT {}\ \BBA {} Belkhir, K.%
\end{APACrefauthors}%
\unskip\
\newblock
\APACrefYearMonthDay{2001}{}{}.
\newblock
{\BBOQ}\APACrefatitle {A {Bayesian} approach to the identification of panmictic
  populations and the assignment of individuals} {A {Bayesian} approach to the
  identification of panmictic populations and the assignment of
  individuals}.{\BBCQ}
\newblock
\APACjournalVolNumPages{Genetics Research}{78}{1}{59--77}.
\PrintBackRefs{\CurrentBib}

\bibitem [\protect \citeauthoryear {%
DeGroot%
}{%
DeGroot%
}{%
{\protect \APACyear {1970}}%
}]{%
degroot1970}
\APACinsertmetastar {%
degroot1970}%
\begin{APACrefauthors}%
DeGroot, M\BPBI H.%
\end{APACrefauthors}%
\unskip\
\newblock
\APACrefYear{1970}.
\newblock
\APACrefbtitle {Optimal Statistical Decisions} {Optimal statistical decisions}.
\newblock
\APACaddressPublisher{}{McGraw-Hill}.
\PrintBackRefs{\CurrentBib}

\bibitem [\protect \citeauthoryear {%
Fomin%
\ \BBA {} Kratsch%
}{%
Fomin%
\ \BBA {} Kratsch%
}{%
{\protect \APACyear {2010}}%
}]{%
fomin2010}
\APACinsertmetastar {%
fomin2010}%
\begin{APACrefauthors}%
Fomin, F.%
\BCBT {}\ \BBA {} Kratsch, D.%
\end{APACrefauthors}%
\unskip\
\newblock
\APACrefYear{2010}.
\newblock
\APACrefbtitle {Exact exponential algorithms} {Exact exponential algorithms}.
\newblock
\APACaddressPublisher{}{Springer-Verlag}.
\PrintBackRefs{\CurrentBib}

\bibitem [\protect \citeauthoryear {%
GMP}{%
GMP}{%
{\protect \APACyear {{\protect \bibnodate {}}}}%
}]{%
GMP}
\APACinsertmetastar {%
GMP}%
\APACrefbtitle {{GMP} -- {The} {GNU} multiple precision arithmetic library.}
  {{GMP} -- {The} {GNU} multiple precision arithmetic library.}
\newblock
\APACrefYearMonthDay{{\protect \bibnodate {}}}{}{}.
\newblock
\APAChowpublished {http://gmplib.org/}.
\PrintBackRefs{\CurrentBib}

\bibitem [\protect \citeauthoryear {%
Hartigan%
}{%
Hartigan%
}{%
{\protect \APACyear {1990}}%
}]{%
hartigan1990}
\APACinsertmetastar {%
hartigan1990}%
\begin{APACrefauthors}%
Hartigan, J.%
\end{APACrefauthors}%
\unskip\
\newblock
\APACrefYearMonthDay{1990}{}{}.
\newblock
{\BBOQ}\APACrefatitle {Partition models} {Partition models}.{\BBCQ}
\newblock
\APACjournalVolNumPages{Communications in Statistics -- Theory and
  Methods}{19}{8}{2745--2756}.
\PrintBackRefs{\CurrentBib}

\bibitem [\protect \citeauthoryear {%
Hubert%
, Arabie%
\BCBL {}\ \BBA {} Meulman%
}{%
Hubert%
\ \protect \BOthers {.}}{%
{\protect \APACyear {2001}}%
}]{%
hubert2001}
\APACinsertmetastar {%
hubert2001}%
\begin{APACrefauthors}%
Hubert, L\BPBI J.%
, Arabie, P.%
\BCBL {}\ \BBA {} Meulman, J\BPBI J.%
\end{APACrefauthors}%
\unskip\
\newblock
\APACrefYear{2001}.
\newblock
\APACrefbtitle {Combinatorial Data Analysis: Optimization by Dynamic
  Programming} {Combinatorial data analysis: Optimization by dynamic
  programming}.
\newblock
\APACaddressPublisher{}{SIAM}.
\PrintBackRefs{\CurrentBib}

\bibitem [\protect \citeauthoryear {%
Huelsenbeck%
\ \BBA {} Andolfatto%
}{%
Huelsenbeck%
\ \BBA {} Andolfatto%
}{%
{\protect \APACyear {2007}}%
}]{%
huelsenbeck2007}
\APACinsertmetastar {%
huelsenbeck2007}%
\begin{APACrefauthors}%
Huelsenbeck, J\BPBI P.%
\BCBT {}\ \BBA {} Andolfatto, P.%
\end{APACrefauthors}%
\unskip\
\newblock
\APACrefYearMonthDay{2007}{}{}.
\newblock
{\BBOQ}\APACrefatitle {Inference of Population Structure Under a {Dirichlet}
  Process Model} {Inference of population structure under a {Dirichlet} process
  model}.{\BBCQ}
\newblock
\APACjournalVolNumPages{Genetics}{175}{4}{1787--1802}.
\PrintBackRefs{\CurrentBib}

\bibitem [\protect \citeauthoryear {%
Jain%
\ \BBA {} Neal%
}{%
Jain%
\ \BBA {} Neal%
}{%
{\protect \APACyear {2004}}%
}]{%
jain2004}
\APACinsertmetastar {%
jain2004}%
\begin{APACrefauthors}%
Jain, S.%
\BCBT {}\ \BBA {} Neal, R.%
\end{APACrefauthors}%
\unskip\
\newblock
\APACrefYearMonthDay{2004}{}{}.
\newblock
{\BBOQ}\APACrefatitle {A split-merge {Markov} chain {Monte Carlo} procedure for
  the {Dirichlet} process mixture model} {A split-merge {Markov} chain {Monte
  Carlo} procedure for the {Dirichlet} process mixture model}.{\BBCQ}
\newblock
\APACjournalVolNumPages{Journal of Computational and Graphical
  Statistics}{13}{1}{158--182}.
\PrintBackRefs{\CurrentBib}

\bibitem [\protect \citeauthoryear {%
Jensen%
}{%
Jensen%
}{%
{\protect \APACyear {1969}}%
}]{%
jensen1969}
\APACinsertmetastar {%
jensen1969}%
\begin{APACrefauthors}%
Jensen, R\BPBI E.%
\end{APACrefauthors}%
\unskip\
\newblock
\APACrefYearMonthDay{1969}{}{}.
\newblock
{\BBOQ}\APACrefatitle {A dynamic programming algorithm for cluster analysis} {A
  dynamic programming algorithm for cluster analysis}.{\BBCQ}
\newblock
\APACjournalVolNumPages{Journal of the Operations Research Society of
  America}{17}{6}{1034--1057}.
\PrintBackRefs{\CurrentBib}

\bibitem [\protect \citeauthoryear {%
Knorr-Held%
\ \BBA {} Ra{\ss}er%
}{%
Knorr-Held%
\ \BBA {} Ra{\ss}er%
}{%
{\protect \APACyear {2000}}%
}]{%
knorrheld2000}
\APACinsertmetastar {%
knorrheld2000}%
\begin{APACrefauthors}%
Knorr-Held, L.%
\BCBT {}\ \BBA {} Ra{\ss}er, G.%
\end{APACrefauthors}%
\unskip\
\newblock
\APACrefYearMonthDay{2000}{}{}.
\newblock
{\BBOQ}\APACrefatitle {Bayesian Detection of Clusters and Discontinuities in
  Disease Maps} {Bayesian detection of clusters and discontinuities in disease
  maps}.{\BBCQ}
\newblock
\APACjournalVolNumPages{Biometrics}{56}{1}{13--21}.
\PrintBackRefs{\CurrentBib}

\bibitem [\protect \citeauthoryear {%
Lau%
\ \BBA {} Green%
}{%
Lau%
\ \BBA {} Green%
}{%
{\protect \APACyear {2007}}%
}]{%
lau2007}
\APACinsertmetastar {%
lau2007}%
\begin{APACrefauthors}%
Lau, J\BPBI W.%
\BCBT {}\ \BBA {} Green, P\BPBI J.%
\end{APACrefauthors}%
\unskip\
\newblock
\APACrefYearMonthDay{2007}{}{}.
\newblock
{\BBOQ}\APACrefatitle {Bayesian model-based clustering procedures} {Bayesian
  model-based clustering procedures}.{\BBCQ}
\newblock
\APACjournalVolNumPages{Journal of Computational and Graphical
  Statistics}{16}{}{526--558}.
\PrintBackRefs{\CurrentBib}

\bibitem [\protect \citeauthoryear {%
Murphy%
}{%
Murphy%
}{%
{\protect \APACyear {2007}}%
}]{%
murphy2007}
\APACinsertmetastar {%
murphy2007}%
\begin{APACrefauthors}%
Murphy, K\BPBI P.%
\end{APACrefauthors}%
\unskip\
\newblock
\APACrefYearMonthDay{2007}{}{}.
\newblock
\APACrefbtitle {Conjugate {Bayesian} analysis of the {Gaussian} distribution}
  {Conjugate {Bayesian} analysis of the {Gaussian} distribution}\
  \APACbVolEdTR{}{\BTR{}}.
\newblock
\APACrefnote{Available at
  \url{http://www.cs.ubc.ca/~murphyk/Papers/bayesGauss.pdf}}
\PrintBackRefs{\CurrentBib}

\bibitem [\protect \citeauthoryear {%
Neal%
}{%
Neal%
}{%
{\protect \APACyear {2000}}%
}]{%
neal2000}
\APACinsertmetastar {%
neal2000}%
\begin{APACrefauthors}%
Neal, R.%
\end{APACrefauthors}%
\unskip\
\newblock
\APACrefYearMonthDay{2000}{}{}.
\newblock
{\BBOQ}\APACrefatitle {Markov chain sampling methods for {Dirichlet} process
  mixture models} {Markov chain sampling methods for {Dirichlet} process
  mixture models}.{\BBCQ}
\newblock
\APACjournalVolNumPages{Journal of Computational and Graphical
  Statistics}{9}{2}{249--265}.
\PrintBackRefs{\CurrentBib}

\bibitem [\protect \citeauthoryear {%
O'Hagan%
}{%
O'Hagan%
}{%
{\protect \APACyear {1997}}%
}]{%
ohagan1997}
\APACinsertmetastar {%
ohagan1997}%
\begin{APACrefauthors}%
O'Hagan, A.%
\end{APACrefauthors}%
\unskip\
\newblock
\APACrefYearMonthDay{1997}{}{}.
\newblock
{\BBOQ}\APACrefatitle {Contribution to discussion of '{On} {Bayesian} analysis
  of mixtures with an unknown number of components' by {S. Richardson} and {P.
  J. Green}} {Contribution to discussion of '{On} {Bayesian} analysis of
  mixtures with an unknown number of components' by {S. Richardson} and {P. J.
  Green}}.{\BBCQ}
\newblock
\APACjournalVolNumPages{Journal of the Royal Statistical Society
  B}{59}{4}{772}.
\PrintBackRefs{\CurrentBib}

\bibitem [\protect \citeauthoryear {%
Quintana%
\ \BBA {} Iglesias%
}{%
Quintana%
\ \BBA {} Iglesias%
}{%
{\protect \APACyear {2003}}%
}]{%
quintana2003}
\APACinsertmetastar {%
quintana2003}%
\begin{APACrefauthors}%
Quintana, F\BPBI A.%
\BCBT {}\ \BBA {} Iglesias, P\BPBI L.%
\end{APACrefauthors}%
\unskip\
\newblock
\APACrefYearMonthDay{2003}{}{}.
\newblock
{\BBOQ}\APACrefatitle {Bayesian clustering and product partition models}
  {Bayesian clustering and product partition models}.{\BBCQ}
\newblock
\APACjournalVolNumPages{Journal of the Royal Statistical Society
  B}{65}{2}{557--574}.
\PrintBackRefs{\CurrentBib}

\bibitem [\protect \citeauthoryear {%
Stephens%
}{%
Stephens%
}{%
{\protect \APACyear {2000}}%
}]{%
stephens2000}
\APACinsertmetastar {%
stephens2000}%
\begin{APACrefauthors}%
Stephens, M.%
\end{APACrefauthors}%
\unskip\
\newblock
\APACrefYearMonthDay{2000}{}{}.
\newblock
{\BBOQ}\APACrefatitle {Dealing with label switching in mixture models} {Dealing
  with label switching in mixture models}.{\BBCQ}
\newblock
\APACjournalVolNumPages{Journal of the Royal Statistical Society
  B}{62}{4}{795--809}.
\PrintBackRefs{\CurrentBib}

\bibitem [\protect \citeauthoryear {%
van Os%
\ \BBA {} Meulman%
}{%
van Os%
\ \BBA {} Meulman%
}{%
{\protect \APACyear {2004}}%
}]{%
os2004}
\APACinsertmetastar {%
os2004}%
\begin{APACrefauthors}%
van Os, B\BPBI J.%
\BCBT {}\ \BBA {} Meulman, J\BPBI J.%
\end{APACrefauthors}%
\unskip\
\newblock
\APACrefYearMonthDay{2004}{}{}.
\newblock
{\BBOQ}\APACrefatitle {Improving dynamic programming strategies for
  partitioning} {Improving dynamic programming strategies for
  partitioning}.{\BBCQ}
\newblock
\APACjournalVolNumPages{Journal of Classification}{21}{2}{207--230}.
\PrintBackRefs{\CurrentBib}

\end{thebibliography}

\end{document}